\documentclass{PoS}
\newcommand{\pf}[1]{{#1 \text{pf}}}
\usepackage{amsmath}
\usepackage{url}

\title{Status and challenges of simulations with dynamical fermions }

\ShortTitle{Status and challenges of simulations with dynamical fermions }

\author{\speaker{Stefan Schaefer}\\
	CERN, Physics Department, 1211 Geneva 23, Switzerland \\ 
        E-mail: \email{stefan.schaefer@cern.ch}}

\abstract{An overview over the current state of algorithms for 
dynamical fermion simulations is given. In particular some insight
into the functioning of the determinant spitting techniques is
discussed. The critical slowing down of the simulations towards
the continuum limit and the role of the boundary conditions is
also reviewed.\\
\mbox{}\hfill{\tt CERN-PH-TH-2012-322}
}

\FullConference{The 30th International Symposium on Lattice Field Theory\\
		 June 24--29,  2012\\
		 Cairns, Australia}

\begin{document}

\section{Introduction}
Continuous work on algorithms has improved our ability to perform
simulations of QCD at light quark masses and on fine lattices. This
was possible due to progress in many domains: solvers for the
Dirac equation with only a mild scaling towards the chiral limit 
have been developed as well as better representations of the quark
determinant that
allow larger step sizes in the integration of the molecular dynamics equations
of motion. These methods are combined with reweighting techniques to
stabilize the simulations and many other  improvements concerning
the numerical and theoretical aspects of a simulation. At the same time,
    the computer hardware available to these computations has become
    much more powerful too.

Understanding these  developments and bringing them together in
efficient packages
has reduced the cost of current light quark algorithms 
over the methods available roughly a decade ago by orders of magnitude.
A review of some of these methods and an approach towards a theoretical
understanding of the determinant splitting techniques is presented 
in the first part of this contribution.

The focus of these developments has been the ability to simulate light
quarks, but for controlled results, also taking the continuum limit is
essential. This has so far been hindered by the problem of the freezing 
topological charge as the lattice spacing is decreased. Practically, this
makes large volume simulations below $0.05\,\text{fm}$ exceedingly difficult,
because below this value,  the autocorrelations associated with the 
topological charge become much longer than typical run lengths and 
reliable measurements are therefore impossible.

A  solution to this problem is to choose a setup where topological
sectors do not form in the continuum limit, by using open boundary
conditions in time. With these boundary conditions, the topological
observables show, in numerical simulations with pure gauge theory,
precisely the scaling of the autocorrelation times as
expected from the algorithm, i.e., the $\tau_\text{int}$
rise with the inverse lattice spacing squared.  This then
provides a basis for the estimation of the required run
length as the simulations move towards the continuum limit,
as will be explained in the second part of this writeup.

\section{Update algorithms\label{s:1}}
Practically all current simulations with dynamical fermions are performed
using a variant of the Hybrid Monte Carlo algorithm
(HMC)\cite{Duane:1987de}. Here the gauge field update is achieved
by introducing momenta $\pi$ conjugate to the field variables and
numerically integrating the molecular dynamics (MD) equations of motion
derived from the Hamiltonian
\begin{equation}
H[\pi,U]=\frac{1}{2}(\pi,\pi)+S_g[U]+S_f[U] \, ,
\end{equation}
where $S_g[U]$ and $S_f[U]$ give the gauge and fermion part of the
action. 

The numerical integrators for the
molecular dynamics equations of motion are not exact,
but lead to an energy violation $\delta H=H_2-H_1$ after a certain
molecular dynamics time $\tau$, with $H_1$ and $H_2$ being,
          respectively,  the value of
$H$ at
the beginning and the end of the trajectory. 
Despite this integration
error, the algorithm is made exact with a Metropolis step, where
the configuration is accepted with probability
\begin{equation}
P_\text{acc}=\text{min}\{1,\exp(-\delta H)\} \,.
\end{equation}

It turns out that a higher acceptance rate can not only
be reached with a smaller step size and by tuning the integrator 
of the MD equations of motion, but also by using the flexibility
in the representation of the fermion determinant. It is in particular 
this choice of $S_f$, or more precisely the factorization of the fermion
matrix, where the various fermion algorithms, like Hasenbusch's mass splitting,
the RHMC or the DD-HMC, differ.

As will become clear in the following,
the crucial point is that  one cannot  discuss the merits of
different effective fermion actions on their own, without taking
into account the interplay between the action
and the integrator. We therefore first introduce the integration
algorithms in Sec.~\ref{s:int} and how to analyze their performance using the shadow
Hamiltonian in Sec.~\ref{s:tune}. In this framework, we
can then analyze the effect of the determinant splitting
(Sec.~\ref{s:split}), which
we demonstrate for the Hasenbusch decomposition in Sec.~\ref{s:hb}.

\subsection{Integration algorithms\label{s:int}}
The HMC algorithm requires the use of symplectic and reversible integrators, 
which are conveniently constructed by alternating updates
of the fields $T_U$ and the momenta $T_\pi$. 
A popular second order integrator\cite{Omelyan2003272}  is given by
\begin{align}
I &= \{T_U(\delta \tau \lambda)\, T_\pi(\delta \tau/2)\, T_U(\delta \tau
         (1-2 \lambda))\, T_\pi(\delta \tau/2)\, T_U(\delta \tau
            \lambda) \}^{n} 
\end{align}
with 
\begin{align}
    T_\pi &: (\pi, U) \to (\pi',U')=(\pi-F(U)\delta \tau,U)
\label{e:int}\\
   T_U &: (\pi, U) \to (\pi',U')=(\pi,U+\pi\,\delta \tau)\,,  \nonumber
   \end{align}
with step size $\delta \tau=\tau/n$ and a tunable parameter $\lambda$. Here we only give a single
time-scale and a single force $F_{x,\mu}^a=\partial_{x,\mu}^aS$, but 
multiple time scale integrators are frequently used\cite{Sexton:1992nu}
and will be briefly discussed in Sec.~\ref{s:mts}. 

The many force
components and integrator options can lead to a bewildering
number of choices.
Theoretical understanding that can facilitate the analysis
of this optimization problem comes from the 
theory of symplectic integrators. For a given Hamiltonian,
the integrator conserves a so-called shadow Hamiltonian, 
which can be constructed as a power series in the step-size $\delta \tau$.
Even though the radius of convergence of this series is not
clear, in practical applications the first order of this
series turns out to give a good approximation at least
for reasonably small step sizes,
as has been shown in a series of papers by Clark  et al., whose
line of argumentation we follow\cite{Clark:2008gh,Clark:2010qw,Clark:2011ir,Kennedy:2012gk}.

For the integrator in Eq.~\ref{e:int}, the shadow Hamiltonian
up to $\delta \tau^2$ reads
\begin{equation}
\widetilde H = H + \delta \tau^2 \{ c_1 \sum_{x,\mu} \partial^a_{x,\mu} S \partial^a_{x,\mu}
S - c_2 \sum_{\substack{x,\mu\\y,\nu}} \pi_{x,\mu}^a \pi_{y,\nu}^b \partial^a_{x,\mu}
\partial^b_{y,\nu} S \} + \dots \equiv H + \Delta H
\label{e:sh}
\end{equation}
with coefficients $c_1=(6\lambda^2-6\lambda+1)/12$ and $c_2=(1-6\lambda)/24$.
Omelyan, Mrygold and Folk\cite{Omelyan2003272} have computed 
the leading term of the shadow
Hamiltonian for a large number of integration schemes, also 
including higher order and force gradient integrators. For the purpose
of this discussion, however, we restrict ourselves to second order integrators. Without further knowledge
of the two second order contributions to $\Delta H$,
they suggest to choose $\lambda$ such that $c_1^2+c_2^2$ is minimized.
As each of the two coefficients has a significantly smaller magnitude
than for the leapfrog integrator (two steps of which correspond to $\lambda=1/4$), this turns out to be a sensible
choice in a typical QCD setup\cite{Takaishi:2005tz},
giving roughly  a gain of a factor two in step-size.

\subsection{Integrator tuning\label{s:tune}}

To improve on this,
Clark  et al. propose to measure the values
of the leading contributions to the shadow Hamiltonian as found 
in the actual simulation. It turns out, that its distribution 
does not change significantly during a trajectory, such that
an equilibrium measurement on a few configurations is 
meaningful. An important insight gained in their work is that the size
of the individual terms does not matter in the first place. This
observation will also be relevant in the following section when the 
choice of fermion action will be discussed, where frequently  the 
size of the fermion force  (the first $\text{O}(\delta \tau^2)$ 
      term in Eq.~\ref{e:sh}) 
has been a leading guide in the analysis of improvements.

To understand why the size of the forces  is an insufficient basis
of argumentation, one first needs to notice that there are 
cancellations between the two
terms at $\text{O}(\delta \tau^2)$. But the size of $\Delta H$  does not matter either.
What ultimately matters only is the acceptance rate in the HMC, in 
which only the difference  $\delta H$ between the value of $H$ at the 
beginning 
and the end of the trajectory  $\delta H=H_2-H_1$ enters. Since up to
corrections of order $\delta \tau^4$ the shadow Hamiltonian $\tilde H$ is 
conserved, we can rewrite this as
\begin{equation}
\delta H = (H_2-\tilde H_2)-(H_1-\tilde H_1)+\text{O}(\delta \tau^4)=\Delta H_1-\Delta H_2 +
\text{O}(\delta \tau^4).
\end{equation}
Because the mean value $\langle \Delta H\rangle$ drops out in this difference,
the {\sl fluctuations} of $\Delta H$ are the quantity
that determines the acceptance rate $P_\text{acc}$.
In fact, for small energy violation
the acceptance rate of the HMC is given by
\begin{equation}
P_\text{acc}=\text{erfc}(\sqrt{\langle \delta H^2\rangle /8}) \ ,
   \label{e:acc}
\end{equation}
and therefore  minimizing $\langle \delta H^2\rangle$ constitutes
a meaningful criterion according to which algorithm can be
optimized. Assuming that the fluctuations $\Delta H_1$ 
and $\Delta H_2$, are independent and equally distributed, this is 
equivalent to minimizing the variance of $\delta H$, because then\cite{Clark:2011ir}
\begin{equation}
\langle \delta H^2\rangle=2\text{var}(\Delta H)\,.
\end{equation}
The acceptance rate is to leading order given by the 
variance of  the {\sl difference} between $H$ and the shadow Hamiltonian.
The value of the norm of the forces drops completely out in this
criterion. Since the variance of the norm squared gives one contribution
to this improvement criterion, it might still be considered in the absence of a measurement
of the second derivative of the action.

\subsection{\label{s:det}Fermion determinant}
The strategy to evaluate integrator improvement can now be used
to provide some understanding in the functioning of the 
determinant splitting techniques which have brought such 
dramatic progress towards realistic fermion simulations.

In the standard HMC, the  quark matrix determinant --- here for simplicity
for two degenerate flavors --- is represented using a single
pseudofermion\cite{Weingarten:1980hx} field $\phi$
\begin{equation}
\det \, Q^2 =\frac{1}{Z_\phi} \int [\text{d} \phi] [\text{d} \phi^\dagger] \,  \exp \{
-(\phi, Q^{-2} \phi )\}\label{e:1pf}
\end{equation}
with $Q=\gamma_5\,D$ the massive Hermitian Dirac operator and
$\phi$ complex-valued spinor fields. This identity leads to an effective
one pseudofermion action $S_\pf{1}$ which is to be compared to the ``exact''
action $S_\text{ex}$, where
\begin{align}
S_\pf{1}&=(\phi, Q^{-2} \phi )  &&\text{and}&
& S_\text{ex}=-\text{tr}\,\text{log}\, Q^2.
\end{align}
In practice the one pseudofermion action leads to very expensive 
light fermion simulations\cite{Ukawa:2002pc}.

Although the pseudofermion action $S_\pf{1}$ for a given realization
of the field $\phi$ might be very different from $S_\text{ex}$, the force
$F_\pf{1}$ deriving from $S_\pf{1}$ is a 
stochastic estimator of the force $F_\text{ex}$ in the sense that the
average over the pseudofermions of the former equals the latter
\begin{equation}
\langle F_\pf{1}\rangle_\phi =-\text{tr}\, Q^{-2} \delta Q^2= F_\text{ex} \,.
\end{equation}
This is at least true at the beginning of the trajectory.
The fluctuations in this estimator, however, are large such that
$|F_\pf{1}|^2 \gg |F_\text{ex}|^2$ and
\begin{equation}
\langle | F_\pf{1} - \langle F_\pf{1} \rangle_\phi|^2 \rangle_\phi =  \langle |F_\pf{1}|^2
\rangle_\phi-|
F_\text{ex} |^2 \approx  \langle |F_\pf{1}|^2 \rangle_\phi
\,.
\end{equation}
The large values for the norm of the fermions forces observed in
practical simulations are thus a 
consequence of the one pseudofermion force  being a poor stochastic
approximation to $F_\text{ex}$. 

\subsection{Determinant factorizations\label{s:split}}
More suitable representations of the fermion determinant can be found
by splitting the contribution into several parts, each of which is introduced
separately by a pseudofermion field.
The physics motivation for the different methods
varies, some focus on the properties of the stochastic estimator, some
aim at a hierarchy of forces (in size), which then can be integrated
on different time scales --- possibly with the higher frequency forces
being much cheaper to compute. The shadow Hamiltonian analysis can
provide a framework to discuss this in a more systematic manner, 
and it has already become clear that just aiming at smaller forces (or
      equivalently a better stochastic estimate of the determinant) is
not the primary target.\footnote{Smaller forces can still be beneficial
as they help avoid problems in connection with the stability of the MD
   integration.}

Several decompositions have been proposed, the three most popular of
which are mass preconditioning, the RHMC and domain decomposition.
The first of those was introduced by
Hasen\-busch\cite{algo:GHMC} and Hasenbusch and
Jansen\cite{Hasenbusch:2002ai} where the 
determinant is split using a stack of  (twisted) quark masses
$0=\mu_1<\mu_2<\dots<\mu_N$ and
the identity
\begin{equation}
\det Q^2 = \prod_{i=1}^{N-1} \det \frac{Q^2+\mu_i^2}{Q^2+\mu_{i+1}^2} \times
\det ( Q^2+\mu_N^2) \, ,\label{e:hb}
\end{equation}
where each determinant is represented by a single pseudofermion
field.\footnote{There are many versions of this splitting, with
   shifts in the mass, the twisted mass and also applying the
      factorization in Eq.~\ref{e:hb}
   to the  Schur
      complement in even-odd preconditioning. This version is chosen for
      ease of notation.}
These masses can be tuned, a choice which will also influence the
relative cost of their evaluation\cite{algo:urbach}.

Alternatively, the RHMC\cite{algo:RHMC} decomposes the determinant into
equal factors using the $N$-th root of the fermion
matrix, which needs to be implemented by a rational approximation
\begin{equation}
\det Q^2 = \prod_{i=1}^{N} \det \sqrt[N]{Q^2} \ .
\label{e:rhmc}
\end{equation}
Each of these factors is again represented by a pseudofermion. 
The RHMC is primarily used for simulating single flavors,
but is also employed for pairs of degenerate quarks as in
Eq.~\ref{e:rhmc}. The evaluation
of the rational approximation, however, is quite costly, in particular
because the frequently employed multi-shift solvers do not combine
well with the inexact preconditioning techniques used for light fermions.

Finally the DD-HMC algorithm\cite{algo:L2} is based on a geometrical block
decomposition, where the Dirac operator restricted to the blocks
$Q_\Lambda$ is
considered in one factor 
\begin{equation}
\det Q^2 = \prod_\Lambda \det Q_\Lambda^2 \times \det Q_S^2 \ ,
\end{equation}
and the second factor is a matrix which has the same determinant as the Schur 
complement of $Q$ with respect to the block projection. This algorithm 
as been successfully used in many Wilson fermion simulations, however,
it suffers from the links between the blocks which do not get updated
during a trajectory, which causes increased autocorrelation
times\cite{Schaefer:2009xx,Marinkovic:2010eg}.

\begin{figure}
\begin{center}
\includegraphics[width=0.3\textwidth,clip]{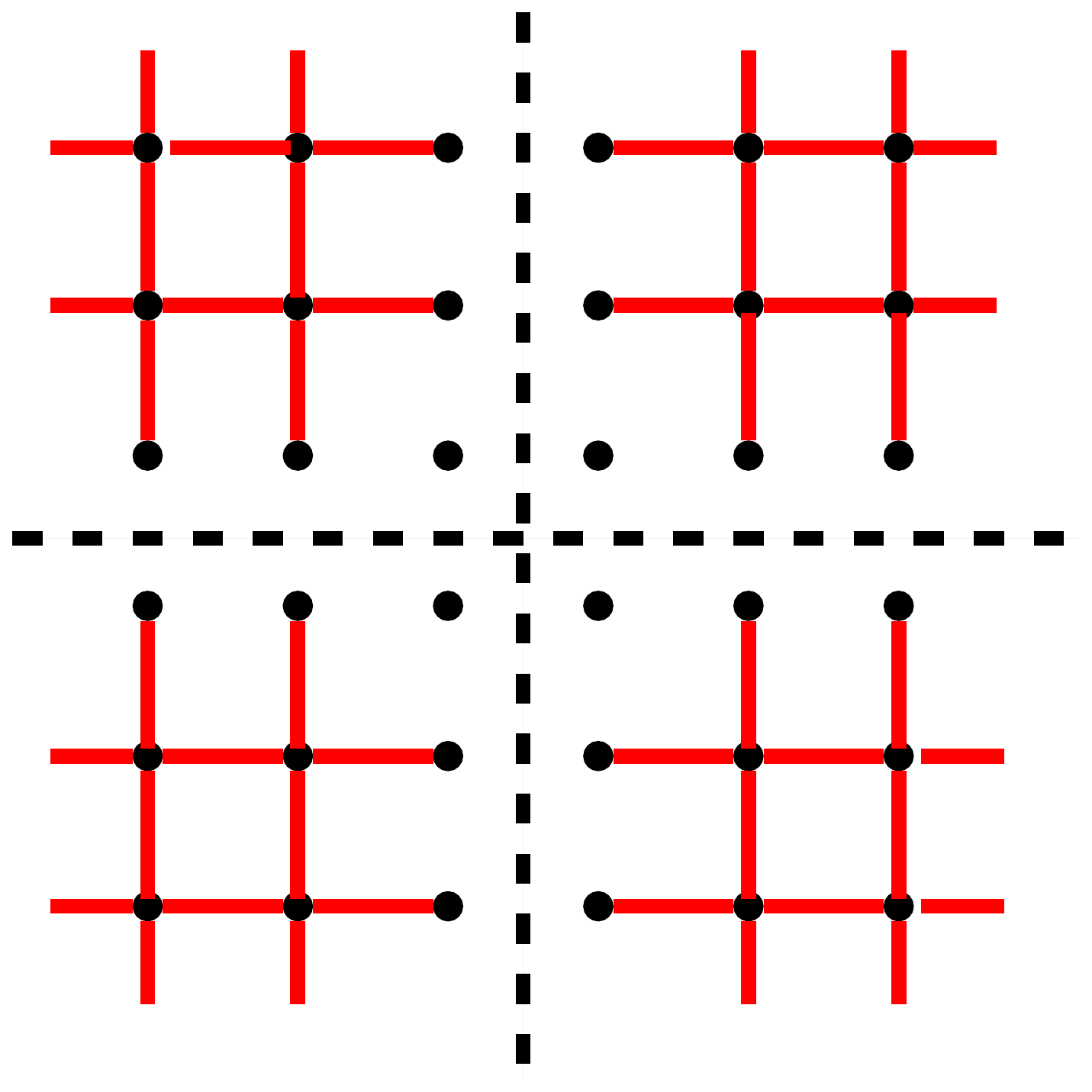}
\end{center}
\caption{\label{f:dd}Decomposition of the Dirac operator in the DD-HMC
   algorithm. The solid lines indicate the gauge links contributing to
      $Q_\Lambda$.}
\end{figure}

A detailed comparison between these algorithms for light quark simulations
has not been published. Just because of its simplicity, we now study
the effect of the  quark determinant splitting at the example of the
Hasenbusch decomposition.

\subsection{Numerical examples\label{s:sim}}

The numerical examples in these proceedings come from a simulation
described in detail in Ref.~\cite{Luscher:2012av}. In particular we
use a run with $2+1$ dynamical flavors of non-perturbatively improved Wilson
fermions and the Iwasaki gauge action. This setup has been extensively
studied by the PACS-CS collaboration\cite{Aoki:2008sm,Aoki:2009ix}.
The pion mass is $200\,\text{MeV}$  and strange quark mass is at
physical value, 
which means that the  $64\times32^3$ lattices at a lattice spacing of
about $a=0.09$\,fm have $L\approx 3.1\,\text{fm}$ and  $m_\pi L\approx 3.1$.

In this simulation, which has been performed with the publicly available
{\tt openQCD} code.\footnote{The code is available under 
\url{http://cern.ch/luscher/openQCD}.} open
boundary conditions in time have been used together with  twisted mass
reweighting, which will both be described below. The three twisted
masses in the frequency splitting, applied on the Schur complement of
the even-odd preconditioning, have been chosen roughly equally
spaced on a log-scale. The strange quark is simulated with the RHMC
algorithm.  Although some details will be different, the main statements
are expected to carry over to different gauge and fermion
discretizations.

\subsection{Effect of the Hasenbusch decomposition\label{s:hb}}

\begin{figure}
\begin{center}
\includegraphics[width=0.4\textwidth,angle=-90,clip]{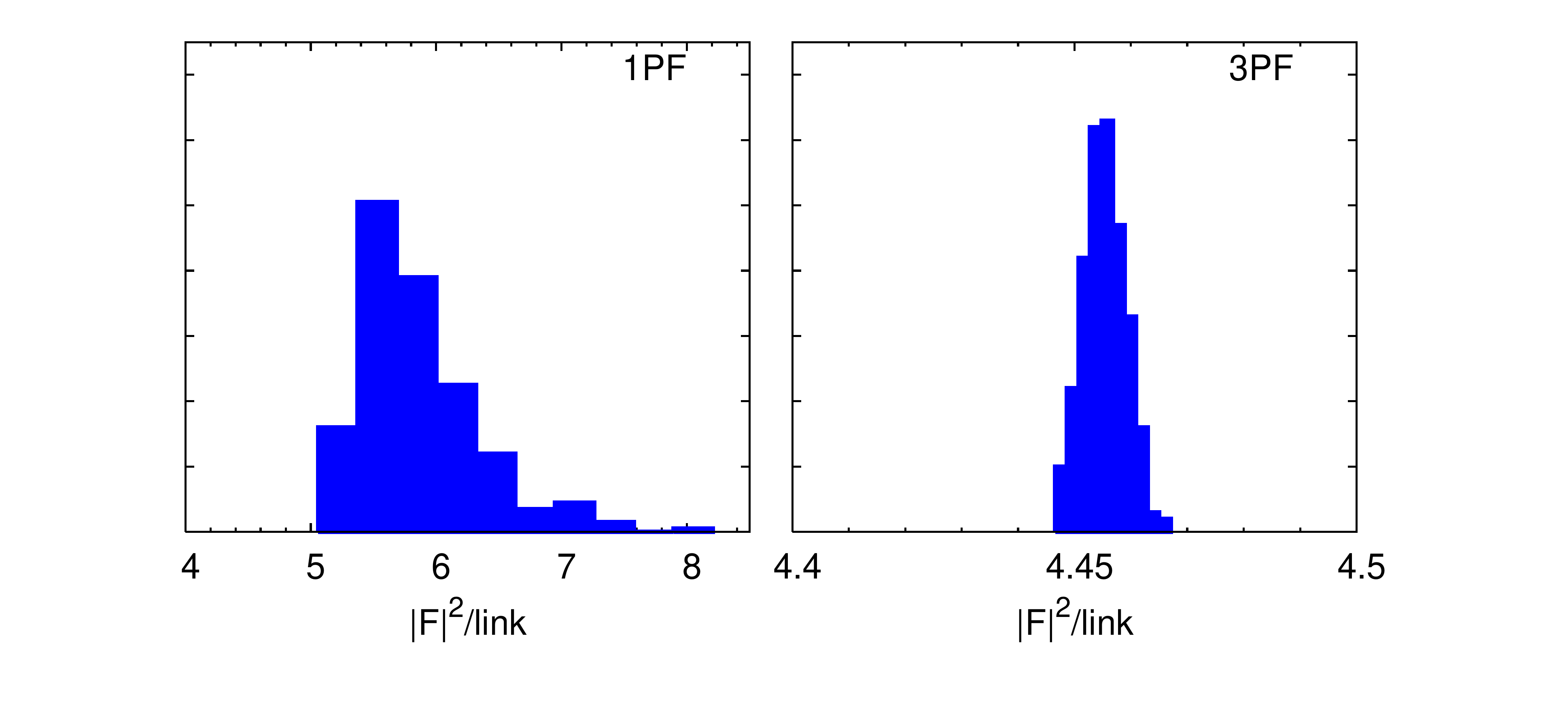}
\end{center}
\caption{\label{f:1}Distribution of the norm of the force with
the one pseudofermion action and using three
pseudofermions in the Hasenbusch splitting. The size of the forces is only marginally reduced, but the fluctuations of the norm of the force decrease drastically. Note the different scale of the $x$-axes.}
\end{figure}

To illustrate the impact of the Hasenbusch mass splitting, we have
measured the first and second derivative of the action on the 
ensemble introduced in the previous section, using $10$ gauge
configurations and $30$ realizations of the pseudofermion fields on each. 
In Fig.~\ref{f:1}, the distribution of the norm squared of the fermion force 
is shown. In the left panel just one pseudofermion field is seen to lead 
to large
fluctuations in the norm of the force. By using the Hasenbusch splitting
with $N=3$, these fluctuations are drastically reduced. Note that the
norm of the force itself is not very much smaller, i.e., the fluctuation
of the force remains large; mass preconditioning is doing little towards
getting a better stochastic estimate of the force.

\begin{figure}
\begin{center}
\includegraphics[width=0.4\textwidth,angle=-90,clip]{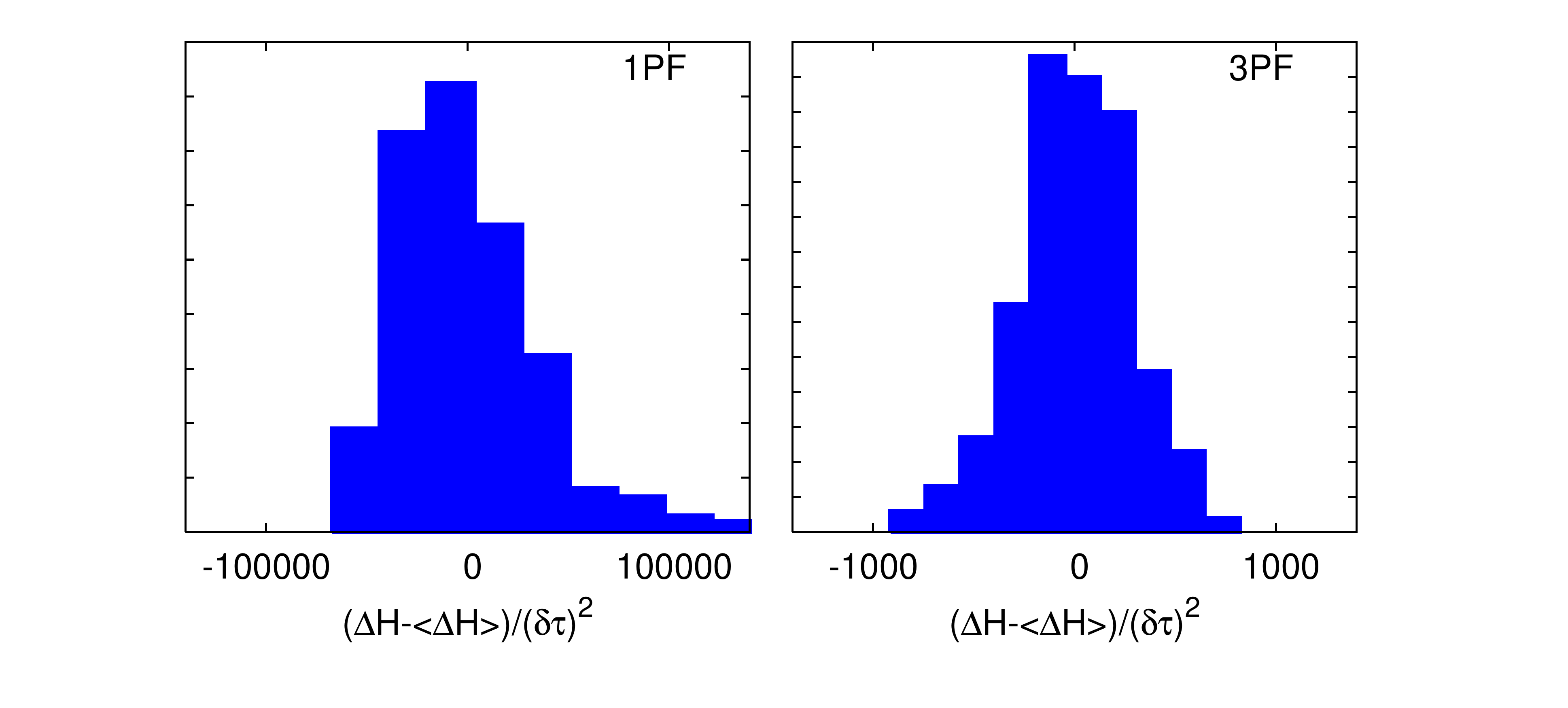}
\end{center}
\caption{\label{f:2} Fluctuation of the value of the shadow Hamiltonian
around its mean value. Left panel
the one pseudofermion action, right panel using three
pseudofermions in the Hasenbusch splitting.}
\end{figure}

However, as we have seen in the previous section, this is not needed.
Smaller fluctuations of the norm of the force are already more significant, but 
what actually counts is the variance of the higher order terms in the
shadow Hamiltonian $\Delta H$ that enters the acceptance rate Eq.~\ref{e:acc}. 
The width of its distribution shown in Fig.~\ref{f:2} is drastically 
reduced by the two
additional pseudofermion fields. To achieve the same acceptance rate,
this allows for an order of magnitude larger step size and a
correspondingly cheaper simulation.

The discussion in this section just wants to illustrate the functioning
of the determinant split-up. Also domain decomposition and the RHMC are
successfully used in modern dynamical fermion simulations, which hints
on a similar effect on the distribution of $\Delta H$ for these
algorithms.

\subsection{Multiple time scales\label{s:mts}}
The size of the gauge and the many components of the fermion
force frequently varies over many orders of magnitude. This
has lead to the suggestion to integrate them on different time scales\cite{Sexton:1992nu}, which is particularly natural if the larger
forces are also much cheaper to compute. The gauge forces, e.g.,
have typically a much larger norm and require significantly
less computer time than the fermion forces. Technically, if
the total action is split in two parts $S=S_1+S_2$, one
integrates one component of the action $S_1$ with the integrator
in eq.~\ref{e:int},
but instead of $T_\pi$, one puts $m$ steps of eq.~\ref{e:int}, where
only the force corresponding to $S_2$ is applied.

Also the split determinants have been combined  with multiple time
scales\cite{AliKhan:2003br,algo:urbach}. From the
shadow Hamiltonian it can be understood why large hierarchies
in the size of the forces do not necessarily translate to corresponding
hierarchies in the time steps.  The shadow Hamiltonian for
the multiple time scale integrator reads\cite{Clark:2008gh}
   \begin{equation}
   \begin{split}
\widetilde H =& H + 
\delta \tau^2 \big\{ c_1 \sum_{x,\mu} \partial^a_{x,\mu} S_1 \partial^a_{x,\mu}
S_1- c_2
\sum_{\substack{x,\mu\\y,\nu}} \pi_{x,\mu}^a \pi_{y,\nu}^b \partial^a_{x,\mu}
\partial^b_{y,\nu} S_1  +c_2\partial^a_{x,\mu} S_1 \partial^a_{x,\mu} S_2 \\ & \frac{1}{m^2} (c_1 \sum_{x,\mu}
      \partial^a_{x,\mu} S_2 \partial^a_{x,\mu}S_2 -  c_2
      \sum_{\substack{x,\mu\\y,\nu}}
      \pi_{x,\mu}^a \pi_{y,\nu}^b \partial^a_{x,\mu} \partial^b_{y,\nu}
      S_2) \big \}  + \dots \,,
   \end{split}
\label{e:sh2}
 \end{equation}
 where the term in the second row reflects the hierarchy of the 
 integration steps. The third term in the first row containing the
 interference between the forces from $S_1$ and $S_2$, however,
 is not suppressed by $1/m^2$ even though it depends on $S_2$. 
 Since in typical settings the forces
 deriving from $S_2$ are much larger than those from $S_1$, this term
 can give a large contribution. 

Obviously, all this depends on the covariance of the various
force contributions and a complete picture needs detailed measurements of
these terms. The smaller the covariance between the two forces, the
better the multiple time scale method will work, but in particular for
the various components of the fermion force, large covariances are to be
expected. In any case,  if the forces from $S_2$ dominate, or are much
cheaper to compute, the multiple time scales can be
beneficial\cite{algo:urbach}. 

To suppress this contribution from the interference term,  the coefficient
$c_2$ can be tuned to zero  by 
choosing $\lambda=1/6$.  Although this is not a good integrator for the
forces deriving from $S_1$ --- it does not profit from the 
cancellation between the two contributions in the shadow Hamiltonian ---
this can be a good choice if these forces and their fluctuations are
small. 

\subsection{Stability of Wilson fermion simulations}
Large forces are known to cause the numerical integrators to become
unstable above a certain step size.  Here light Wilson fermions have a
particular problem due to the lack of a spectral gap, which results in
potentially unbounded forces where eigenvalues become zero.
The most obvious
consequence of this are the so-called ``spikes'', exceptionally large
values of $\delta H$, which are observed during the simulation.
Furthermore, the molecular dynamics trajectories cannot cross the
surfaces of zero eigenvalues, at least if the integration is exact. This
means that the simulation is not ergodic  and ergodicity
depends on integration errors. Also thermalization can be affected by
the evolution being stuck in one of these sectors.   

Introducing a small twisted mass during the simulation into the light
Dirac operator can cure these problems and it has been argued in
Ref.~\cite{Luscher:2008tw} that even on typical large volumes its effect
can be reliably cancelled by including a corresponding reweighting term
into the measurement. Two possible replacements for the fermion
determinant have been proposed
\begin{equation}
\det Q^2 \to \begin{cases}\det (Q^2+\mu^2)  &
   \text{Type I}\\
\det (Q^2+\mu^2)^2/{\det}(Q^2+2\mu^2) & \text{Type II}\,,
\end{cases}
\end{equation}
where the second option has the advantage that the contribution from
large eigenvalues $\lambda$ of $Q$ 
to the reweighting factor $R$ ---corresponding to the ratio of $\det
Q^2$ to what it has been replaced with ---  is suppressed by $\mu^4/\lambda^4$, whereas
the first falls of with $\mu^2/\lambda^2$. Since large contributions in
the ultraviolet limit the reach in $\mu$ of the reweighting, and
therefore the benfits that can be achieved in the infrared, the second
choice is likely to yield better results.

\begin{figure}
\begin{center}
\includegraphics[width=0.4\textwidth,angle=-90,clip]{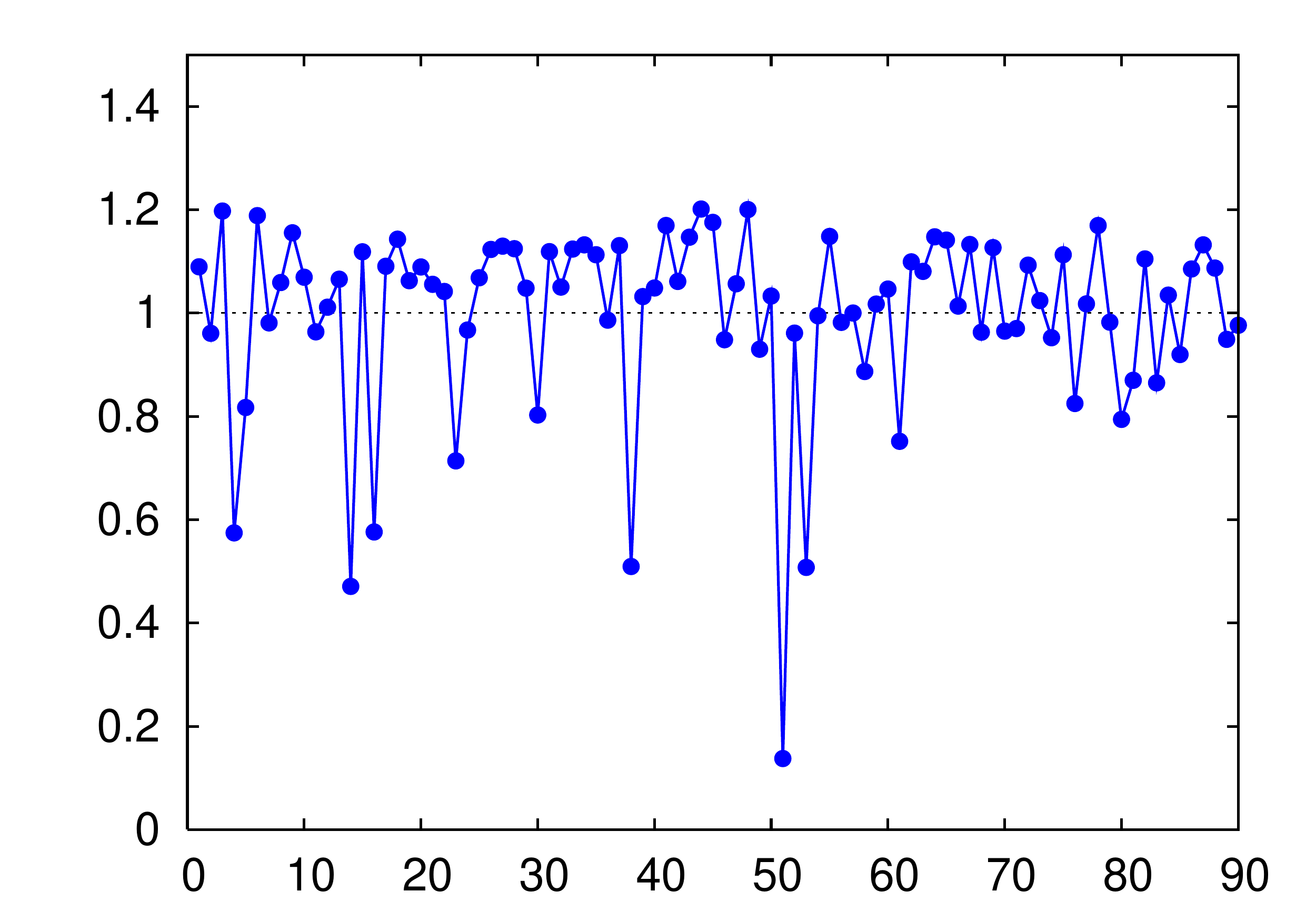}
\end{center}
\caption{\label{f:rw}Reweighting factor $R$ for the second type of
   reweighting in the simulation described in the text.
}
\end{figure}

For the simulation described in Sec.~\ref{s:sim}, the second option
has been employed, with $\mu$ large enough to efficiently suppress
large values of $|\delta H|$. In fact, on at most a few per-mille of the
trajectories it assumed values above $2$.
Still, as can be seen from the measured reweighting factors in Fig.~\ref{f:rw},
its fluctuations are under control and only 15\% of the configurations
obtain a weight below $0.5$. 

Also reweighting in the quark mass (instead of a twisted mass term) has
been proposed\cite{Hasenfratz:2008fg} and makes
part of what is used in large scale
simulations\cite{Aoki:2008sm,Aoki:2009ix,Liu:2012gm}. Here the emphasis is less on
a strict spectral bound but on possible corrections in the tuning or the
possibility to reach smaller quark masses more cheaply. In both
reweighting methods, the over-sampling of small eigenvalues can also
lead to reduced fluctuations in some observables, e.g., the pion
correlator.

\subsection{Solver}
A main difficulty of going towards smaller quark masses has been the
rise of the condition number of the fermion matrix as $am_q\to0$ and the
associated increase in iteration count of the iterative solvers.
Deflation techniques which remove the contribution of the low-modes from
the linear system substantially alter this situation.  Explicitly
computing (an approximation to) the $N$ lowest eigenmodes, however,  is
not an option in large volume simulations, since $N$ needs to be scaled
with $V$ in order to keep the condition number approximately constant.

This has been changed by the advent of local deflation
techniques\cite{algo:coherence}, where the space which is projected from
the system is spanned by local (block) modes. The number of these modes
per block is kept fixed as the volume is increased, which means that the
dimension of the space spanned is growing with the volume. The resulting
iterative, locally deflated solver shows only a very modest, if any, increase in
iteration count as the quark is lowered towards the chiral limit.
Closely related to this approach are adaptive multigrid
algorithms\cite{Babich:2010qb,Frommer:2012mv} which consequently show similar
performance gains and are also being developed for use together with
domain wall fermions\cite{Cohen:2012sh}. 

All these techniques require a certain amount of setup time and also the
memory usage can be substantial. This is, however, easily amortized as
the force evaluation in the  determinant splitting techniques need the
solution of the Dirac equation for many right hand sides on the same
gauge field.


\section{\label{s:top}Continuum limit}
Approaching the continuum limit is an essential part of a
field theory calculation, but in most practical simulations the lattice
spacing can only be varied by a modest factor (and still lie within  the
scaling regime). This is due to the very steep increase in
computational cost as the continuum limit is approached if the physical
volume is to be held fixed.  The number of lattice points increases with
$a^{-4}$ and to get fixed acceptance rate with a second order integrator
requires $a^{-1}$ steps per trajectory. Furthermore,  Monte Carlo time
itself has a scales with the lattice spacing and autocorrelation times of
observables $A$ are expected to show a universal behavior 
\begin{equation}
\tau_\text{int}(A) \propto a^{-z}
\end{equation}
with a dynamical critical exponent $z$, which for the HMC algorithm is
$z\approx2$ as argued in Sec.~\ref{s:scale}.
The total cost
is, thus, expected to rise with the seventh power of the inverse lattice
spacing, each factor of two in $a$ translating into more than two orders
of magnitude in cost.  There might be some effects from reduced noise
due to smoother gauge fields, but it is still necessary to produce a
Monte Carlo time history which contains at least $\text{O}(100)$ times
the largest exponential autocorrelation time, setting a lower bound for
the numerical effort, irrespective of the desired accuracy.

In the traditional setup of QCD simulations using periodic boundary 
conditions for the gauge fields, an additional problem arises due to the 
global topological charge. Field space in the
continuum is disconnected and decomposed into sectors of different integer
topological charge,
\begin{equation}
Q=\frac{1}{32 \pi^2} \int \text{d}x \,\epsilon_{\mu \nu \rho \sigma}\,
F_{\mu \nu} F_{\rho \sigma} \,.
\end{equation}
We therefore expect that close to the continuum limit any algorithm
which changes the fields (quasi)-continuously has difficulty 
changing the global topological charge of the gauge configuration.

\begin{figure}
\begin{center}
\includegraphics[width=0.4\textwidth,angle=-90,clip]{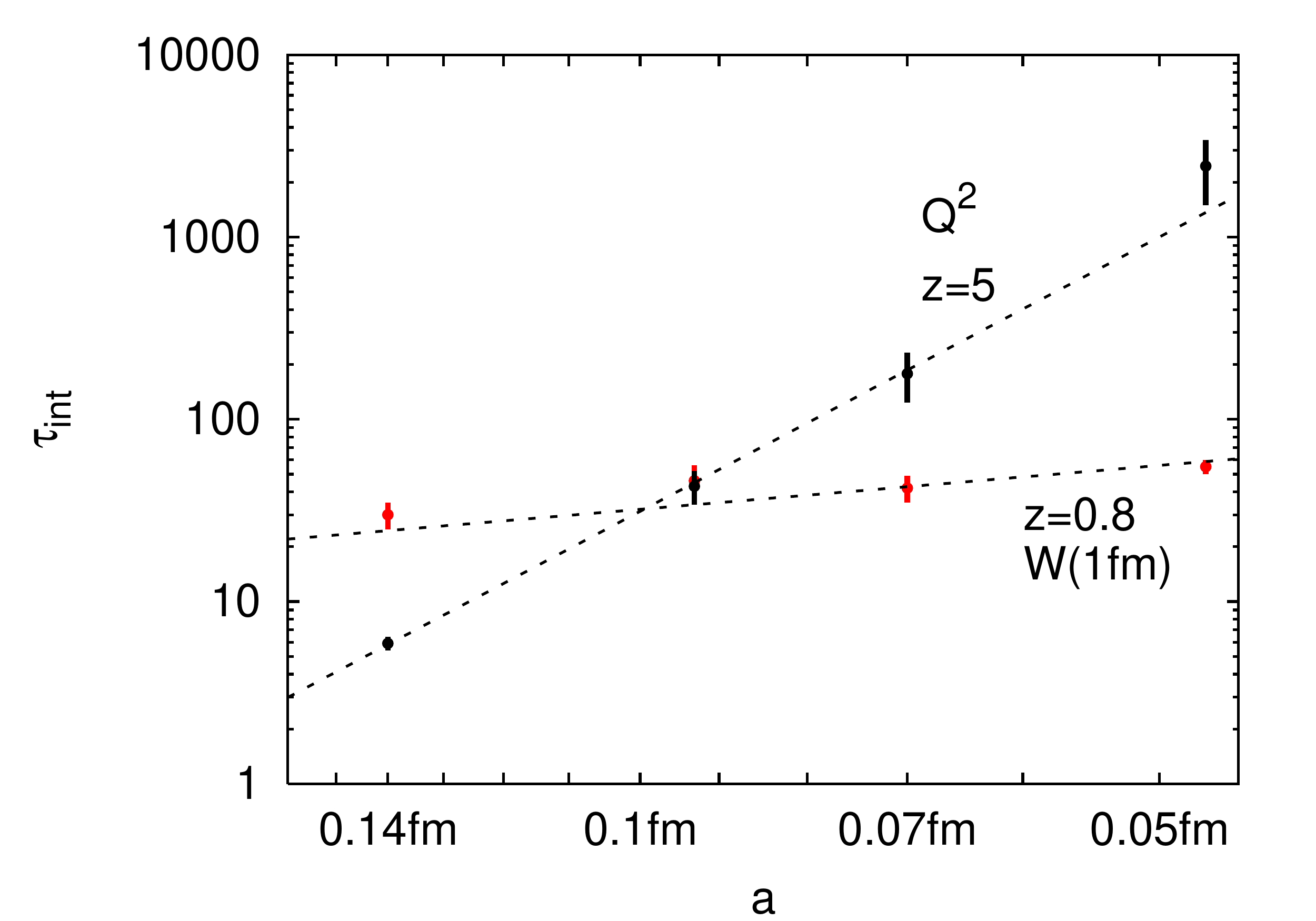}
\end{center}
\caption{\label{f:3}Integrated autocorrelation time vs. lattice spacing
for a smeared Wilson loop of size $(1\text{fm})^2$ and the topological charge in pure gauge
   theory with Wilson gauge action. Data from
      Ref.~\cite{Schaefer:2010hu}.}
\end{figure}

How quickly this continuum behavior
is approached is a priori not clear and details will in general depend
on the gauge and fermion discretizations. However,  pure gauge
theory with  the Wilson gauge action has exhibited a drastic rise in the
autocorrelation time of $Q^2$, compatible with 
$\tau_\text{int}(Q^2) \propto a^{-5}$, see Fig.~\ref{f:3}, but could
also be exponential in $a^{-1}$\cite{DelDebbio:2004xh}.

In a 2010 contribution to this conference series\cite{Luscher:2010we}, 
L\"uscher observed that in pure gauge theory, in a fixed volume the 
probability to find a configuration which has a large plaquette 
value and is therefore in between
topological sectors (plaquettes constructed from links
smoothed by the Wilson flow) decreases as $a^{-6}$ as the continuum
limit is approached and the volume is kept fixed.
This indicates the rapid formation of topological sectors and 
corresponds to the drastic rise in $\tau_\text{int}(Q^2)$ as $a\to0$.

The situation with dynamical fermions has been discussed at this
conference. For the ETM collaboration, Deuzeman\cite{lat12:deuzeman}
reported an absence of  slowing down of the topological charge towards
the continuum, whereas in the Wilson fermion simulations  with the DD-HMC
algorithm  Mondal found a substantial increase in
autocorrelations\cite{Chowdhury:2012qm}. An exceptionally slow evolution
of the topological charge in the presence of dynamical fermions has in
the past also been found for a variety of gauge and fermion
discretizations\cite{Alles:1996vn,Bali:2001gk,Bernard:2003gq,Fritzsch:2012wq}.

Let us briefly note that the problem of a slow global topological charge
could in principle be solved by fixing the topological
sector during the simulation\cite{Fukaya:2006ca}. With a different
motivation, the JLQCD
collaboration has been using this for their dynamical overlap
simulations\cite{Aoki:2008tq}. In this setup
one has to deal with power-like finite volume corrections that vanish
only as $V^{-1}$\cite{topo:fixed1}, and whose analytic form is not known
for all observables. The impact of the fixed topological charge
on the algorithmic performance has so far not been studied close to the
continuum limit.

\subsection{Expected scaling\label{s:scale}}
The question remains whether the observed scaling for the topological
charge is exceptional, or whether there is reason to expect that
autocorrelations should rise with a smaller  power of the inverse lattice
spacing.
From dimensional analysis of the HMC equations of motion
autocorrelation times should scale with the inverse lattice
spacing\cite{Kennedy:2000ju}, once
the trajectory length is scaled accordingly. This does not match with
the experience form measurements as in Fig.~\ref{f:3}.

This free field theory result, however, cannot be expected to hold in the
interacting theory as has been argued in Ref.~\cite{Luscher:2011qa}.
There, an attempt to prove the renormalizability of the five-dimensional
field theory --- defined by the four-dimensional physical field theory
augmented by the dynamics in  simulation time as a fifth dimension ---
has lead to a divergence that could not be removed by a local
counter term. 

These methods have earlier been
used to prove the renormalizability of the Langevin equation for scalar
field theory and gauge theory\cite{ZinnJustin:1986eq,Baulieu:1999wz}.  This
means that up to logarithmic corrections, the free field result of $z=2$
holds for algorithms based on this stochastic differential equation.  In
Ref.~\cite{Luscher:2011qa} it has been  conjectured that the HMC
algorithm falls into the universality class of the Langevin equation.
For observables $A$ with a proper continuum limit, it is therefore
expected that the integrated autocorrelation function scales with the
inverse lattice spacing squared
\begin{align}
\tau_\text{int}(A,W)\, \propto\, a^{-2} &&\text{with}
&&\tau_\text{int}(A,W)=\frac{1}{2}\tau + \tau \sum_{k=1}^W
\frac{\Gamma_A(k \tau)}{\Gamma_A(0)}
\end{align}
and $\Gamma=\langle (A(t)-\bar A)
   (A(0)-\bar A)\rangle$ the autocorrelation function and $\tau$ the distance
   between two measurements.

These arguments involve a continuum limit of the autocorrelation
function and the question for the algorithmic analysis is whether
the limits $a\to 0$ and $W\to \infty$ commute. For the performance of the
algorithms, one is interested at the $W\to\infty$ limit at fixed lattice
spacing, whereas the theoretical arguments apply to taking the continuum
limit first. But in general, a quadratic
scaling should be expected also for $\tau_\text{int}$ and the 
scaling found in the topological charge can indeed be considered to be
exceptionally strong.

\begin{figure}
\begin{center}
\includegraphics[width=0.4\textwidth,angle=-90,clip]{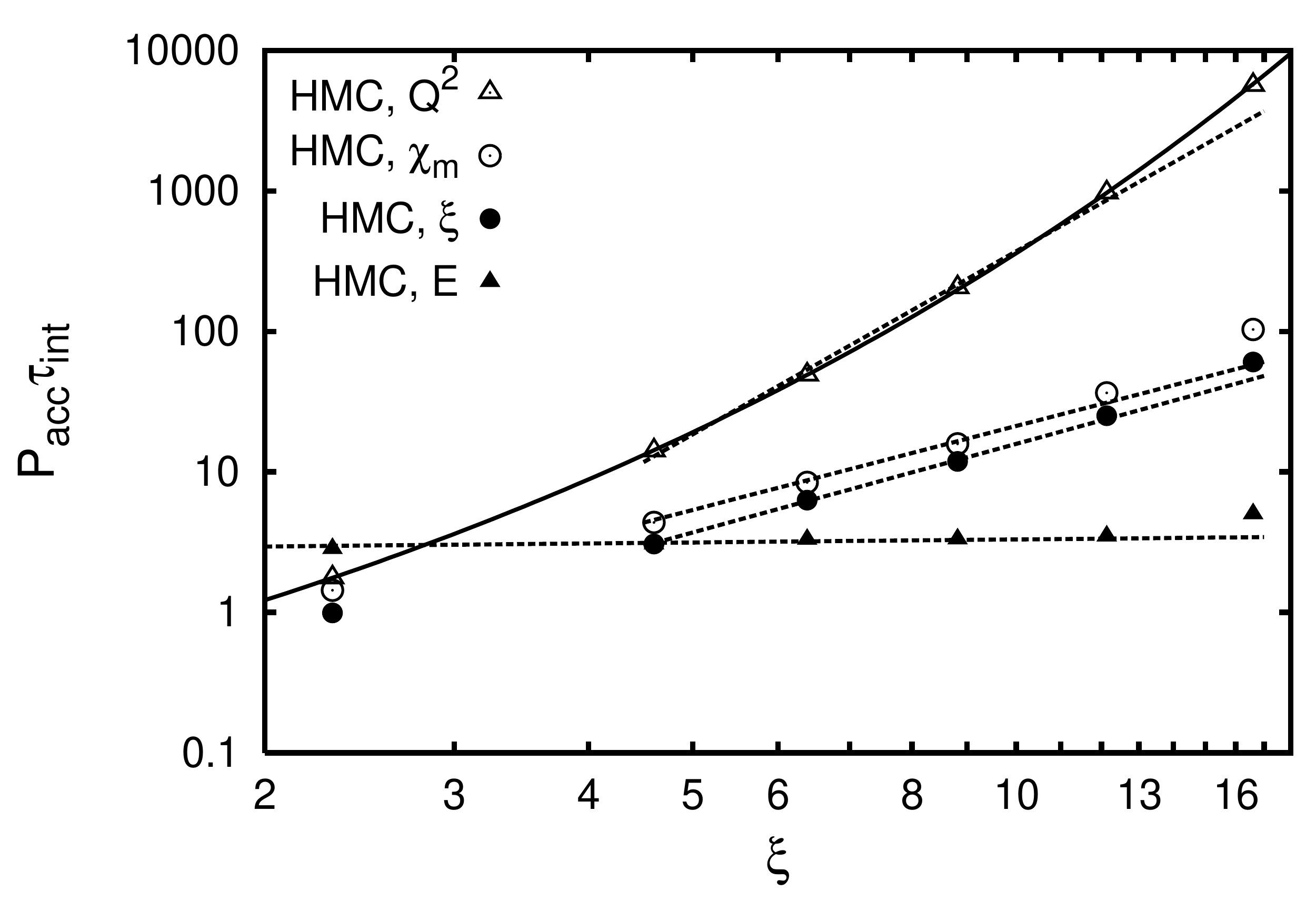}
\end{center}
\caption{\label{f:scaling}Scaling of the integrated autocorrelation
   times with the lattice spacing in the $\text{CP}^9$
      model\cite{Engel:2011re}. The
      topological charge squared shows a behavior compatible with
an exponential growth indicated by the solid line, the dashed line
follows a power law with
$z=4$. For the magnetic susceptibility $\chi_m$ and correlation
length $\xi$ the dashed lines show a scaling with  $\xi^{2}$,
       whereas the corresponding curve for the action density $E$ has
          $z=0.12$.}
\end{figure}

An illustration of these statements can be found in data from the
$\text{CP}^9$ model\cite{Engel:2011re} presented in Fig.~\ref{f:scaling}.
For an earlier discussion of topological slowing down in this model see
Ref.~\cite{DelDebbio:2004xh}.
The magnetic susceptibility
$\chi_m$ and the correlation length $\xi$ show a scaling of
$\tau_\text{int}$ compatible with a dynamical critical exponent $z=2$.
They both have a meaningful continuum limit. This is not the case for 
the action density $E$, which exhibits an almost constant behavior.
However, the topological charge shows a growth in the autocorrelation
time which is compatible with an exponential behavior, as already
proposed in Ref.~\cite{DelDebbio:2004xh}. At a certain point, the
dependence of the other observables on the topological charge --- small
as it may be --- dominates all autocorrelation functions and causes
a deviation from the previous behavior. This also exemplifies the danger
the frozen charge poses for simulations in QCD, where a seemingly
decoupled observable can suddenly receive important contribution to its
autocorrelations from the slow modes.

\subsection{Open boundary conditions}
Since we have argued that exceptionally long autocorrelations
of the topological charge are linked to the topological sectors
in the continuum, a setup in which these sectors do not exist
is likely to solve the problem. In Ref.~\cite{Luscher:2011kk} it
has been proposed to use open boundary conditions {\sl in time}, 
such that the topological charge can continuously change over 
these boundaries. Field space in the continuum is no longer disconnected;
there is no integer topological charge.
In the spatial directions, periodic boundary conditions are kept.
Technically, this means to impose for the
quark and antiquark fields $\psi(x)$ and
$\overline \psi(x)$,
\begin{align}
   \left.P_{+}\psi(x)\right|_{x_0=0}&=\left.P_{-}\psi(x)\right|_{x_0=T}=0,
   & P_{\pm}&=\frac{1}{2}(1\pm\gamma_0)\,,\\
   \left.\overline\psi(x)P_{-}\right|_{x_0=0}&=\left.\overline\psi(x)P_{+}\right|_{x_0=T}=0,\nonumber
\end{align}
like in the Schr\"odinger functional. For the gauge fields
\begin{equation}
      \left.F_{0k}(x)\right|_{x_0=0}=\left.F_{0k}(x)\right|_{x_0=T}=0
            \quad\text{for all}\quad k=1,2,3 
\end{equation}
is chosen.

Since periodic boundary conditions in the spatial directions are imposed, 
the projection to definite momentum of the operators is still possible and 
the impact on the physics analysis is minimal, because in the time direction
the transfer matrix of the theory is not changed. The boundary
conditions will be reflected in the hadronic correlators, but the
particle spectrum will be the same.

\subsection{Observables}
To ensure that a simulation is reliable, it is pivotal to have a Monte
Carlo history that is much longer than the longest exponential
autocorrelation time. To this end it is not sufficient to examine 
the autocorrelation function of the observable in question, because in
a noisy observable the exponential tail in the autocorrelation function 
can easily be covered by
statistical fluctuations; a seemingly small $\tau_\text{int}(A)$ can be a 
consequence of a short run history. For the arguments concerning the
renormalizability of the algorithm to apply, it is also beneficial to
consider quantities with a well defined continuum limit

Such observables, which also have low statistical noise,  can be constructed
using the Wilson flow\cite{Narayanan:2006rf,Luscher:2010iy,Luscher:2011bx}, 
defined by a partial differential equation on the gauge fields
\begin{align} 
\partial_t V_t(x,\mu) &= -ag_0^2 T^a (\partial^a_{x,\mu}S_\text{W})(V_t) V_t(x,\mu)\,,& 
    \left.V_t(x,\mu)\right|_{t=0}&=U(x,\mu) \,.
\end{align}
Starting from the gauge fields $U(x,\mu)$, integrating this equation up to a 
flow time $t$ defines gauge fields $V(x,\mu)$ that are smoothed out over a radius
$r=\sqrt{8\,t}$.

In the following, we will consider three such observables, the action density and
the topological charge density summed over a time slice as well as the global topological charge
\begin{align}
       \overline Q(x_0)&=-\frac{a^3}{32\pi^2}\,\sum_{\vec{x}}
          \epsilon_{\mu\nu\rho\sigma}\,
	     \text{tr}\,\{G_{\mu\nu}(x)\,G_{\rho\sigma}(x)\}\,; & Q&=a\sum_{x_0}\overline Q(x_0)\,;\\
       \overline E(x_0)&=-\frac{a^3}{2L^3}\,\sum_{\vec{x}}
          \text{tr}\,\{G_{\mu\nu}(x)\,G_{\mu\nu}(x)\}\ .\nonumber
\end{align}
For the sliced observables, the central time slice will  be considered.
The crucial point of these observables is that the smoothing radius is kept fixed
as the continuum limit is approached.
In the following, the flow time will be taken to at $t_0$, i.e., the flow 
time at which $t^2 \langle E \rangle|_{t=t_0} = 0.3$ intoduced in
Ref.~\cite{Luscher:2010iy}, which corresponds to a 
smoothing radius of $\sqrt{8t_0}\approx
0.42\,\mathrm{fm}$\cite{Borsanyi:2012zs}.

\begin{figure}
\begin{center}
\includegraphics[width=0.4\textwidth,angle=-90,clip]{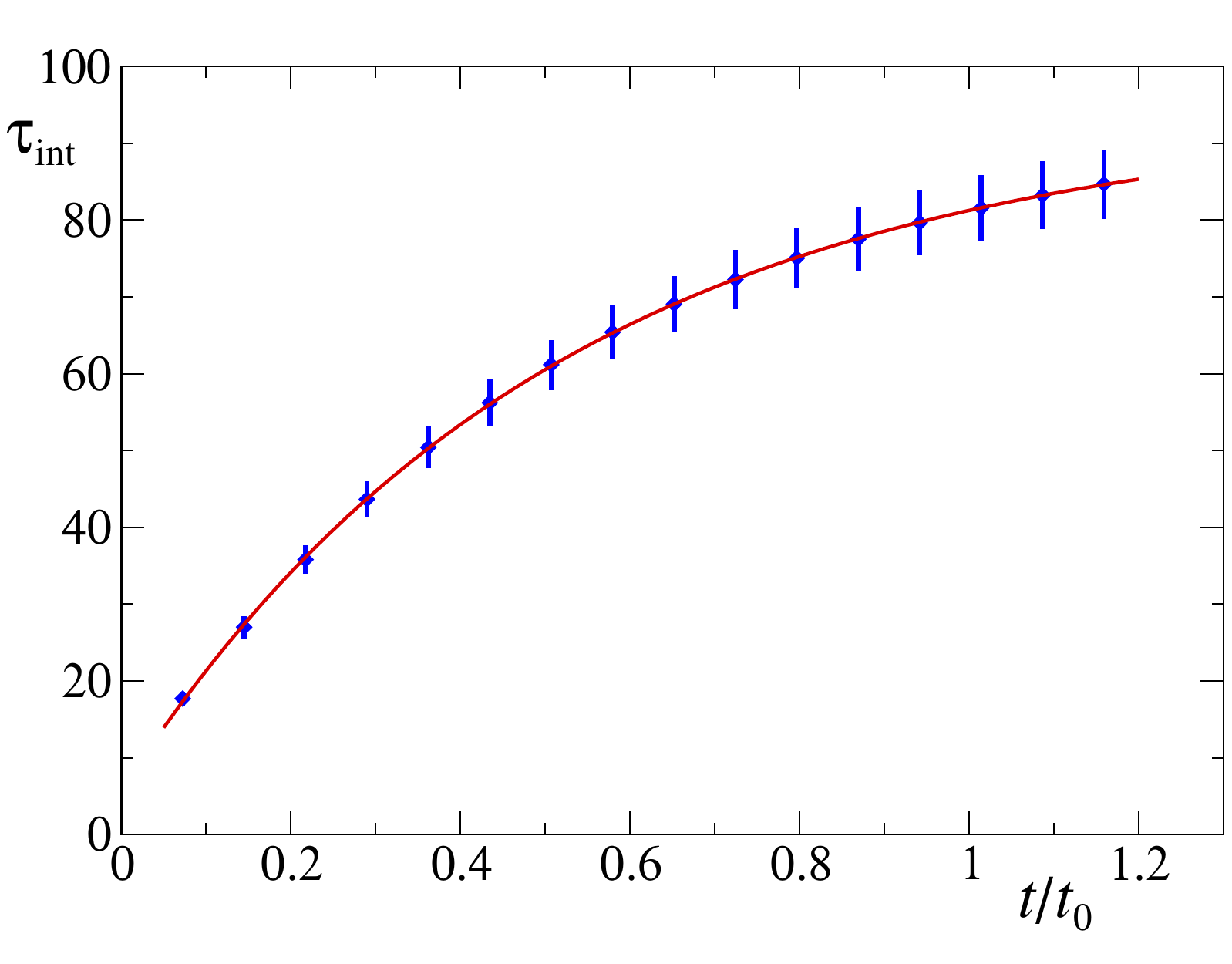}
\end{center}
\caption{\label{f:demo}Integrated autocorrelation time of $\overline E$ as
a function of the flow time in units of $t_0$. The solid line shows a
   fit to the data of the functional dependence discussed in the text.  }
\end{figure}

The smoothing has a notable effect on the integrated autocorrelation
time, as is demonstrated in Fig.~\ref{f:demo} taken from
Ref.~\cite{Luscher:2011kk} for pure gauge theory and
a lattice spacing of $a\approx 0.05\,\text{fm}$. The integrated
autocorrelation time $\tau_\text{int}(\overline E)$ increases by roughly an order of
magnitude, highlighting the importance of choosing smooth quantities
with little noise. The behavior is well described by an exponential
approach to a constant $\tau_\text{int}(\overline E)=a+b \exp(-t/c)$, 
with the autocorrelations at $t=t_0$ being close to saturation.

\subsection{Effect of the open boundary conditions on autocorrelations}

\begin{figure}
\begin{center}
\includegraphics[width=0.8\textwidth,clip]{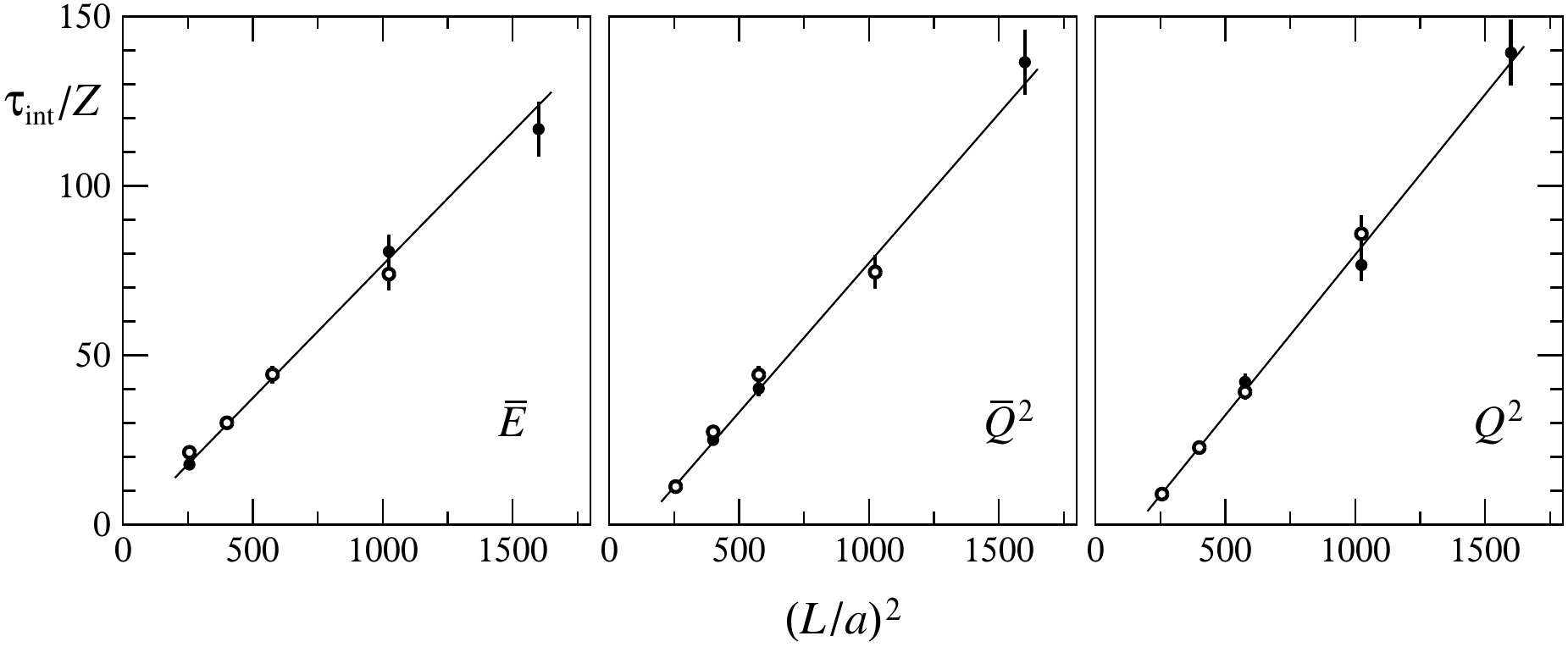}
\end{center}
\caption{\label{f:scalegauge} Scaling of the integrated autocorrelation
times in pure gauge theory with open boundary conditions in time  towards the continuum limit. The observed scaling
matches with a dynamical critical exponent of $z=2$. The topological
charge does not show exceptional slowing down. Filled symbols are
results for the SMD algorithm with $Z=1$, open symbols for the HMC
$Z=1.32$.}
\end{figure}

The open boundary conditions in time have been studied in pure gauge
theory with the Wilson action in Ref.~\cite{Luscher:2011kk} from which 
Fig.~\ref{f:scalegauge} has been taken.  It shows the scaling
of the integrated autocorrelation time with the lattice spacing on
a lattice of constant physical volume $V=(1.6\,\text{fm})^4$.
Results for the stochastic molecular dynamics algorithm(SMD)\cite{Horowitz:1991rr}
and the HMC algorithm are shown. The two data sets show the same scaling
behavior and differ only by a global normalization factor.

All three observables follow the $a^{-2}$ expectation, no sign of topological
freezing could be found, even though the scale covers roughly the same
range of lattice spacings as in Fig.~\ref{f:3}, which was using the same
gauge action with periodic boundary conditions. The autocorrelation time
of the topological charge is significantly reduced by the open
boundaries. For the HMC at $a\approx0.05\,\text{fm}$, we have
$\tau_\text{int}(Q^2)\approx1000$ with the periodic boundary conditions
and $\tau_\text{int}(Q^2)\approx180$ with open boundaries.   Remarkably, the
autocorrelation times of the topological quantities and for $\overline E$
are not very different.

\subsection{Effect on hadron correlation functions}

Since the open boundary conditions work as expected for the
autocorrelations, we now have to study their effect on physical
observables.
On general grounds, one expects their effect
to be exponentially suppressed by the distance from  $x_0=0$ and $x_0=T$.
Closer to the boundaries, of course, the boundary conditions are
reflected in the temporal dependence of the correlators.

\begin{figure}[t]
\begin{center}
\includegraphics[width=0.4\textwidth,angle=-90,clip]{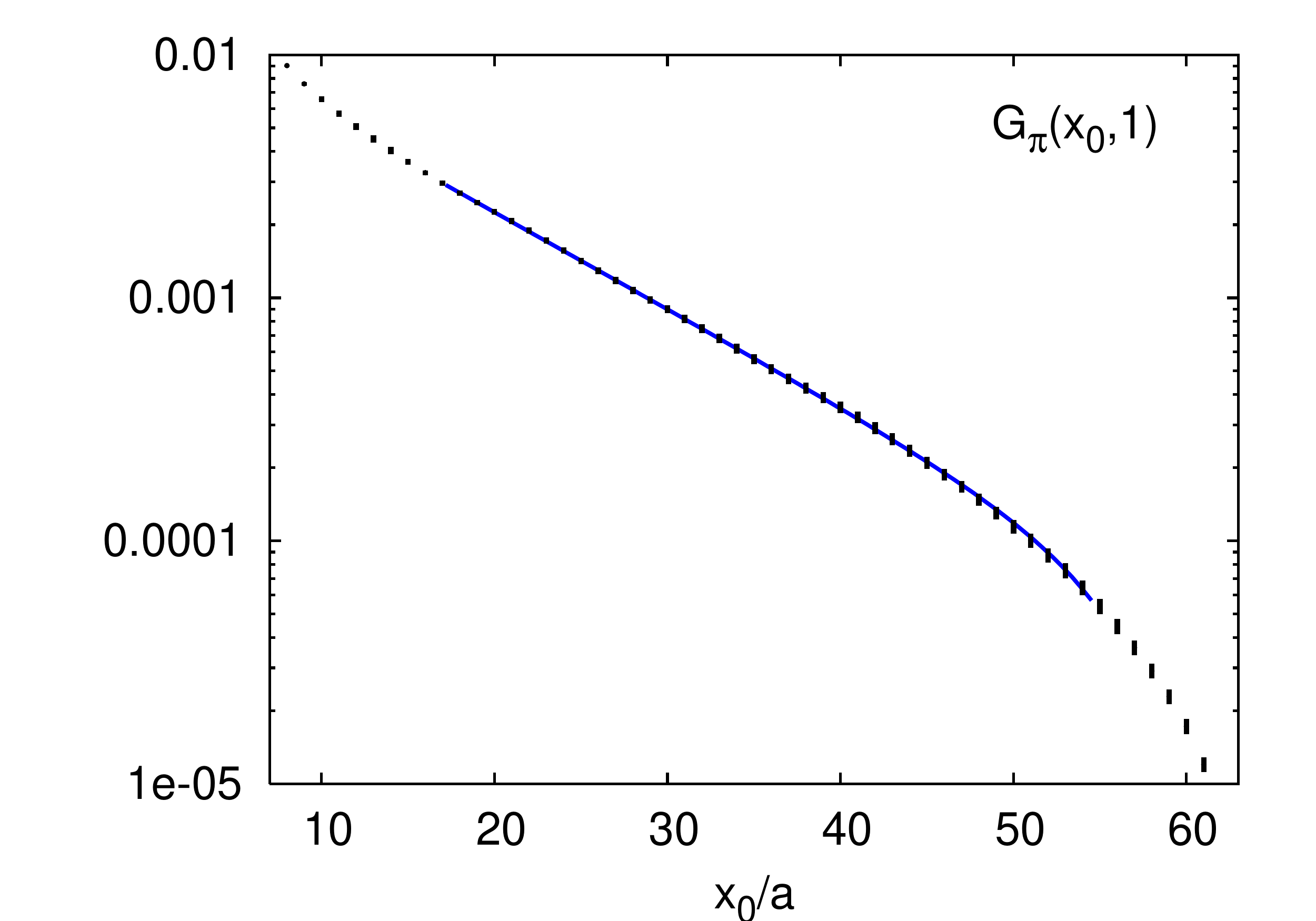}
\end{center}
\caption{\label{f:prop}Pseudoscalar correlation function with the source
at $x_0=a$. A fit to the chiral perturbation theory formula is shown,
which supports the hypothesis of Dirichlet boundary conditions for the
   pion fields.}
\end{figure}

The (light) pseudoscalar correlation function is known to be dominated by pions
in the long distance and light quark mass regime. Since Dirichlet
boundary conditions are generic in scalar field theories, it is
therefore natural to assume that the pion filed $\pi^a$ vanishes on the
boundaries
\begin{equation}
\pi^a(x)|_{x_0=0} = \pi^a(x)|_{x_0=T} = 0 \,.
\end{equation}
For the pion propagator $G_\pi(x_0,y_0)$, one infers by solving the
Klein-Gordon equation that for $x_0 > y_0$ it assumes in the region of
$x_0$ not
too close to the boundaries the form
\begin{equation}
G_\pi(x_0,y_0)\propto  \sinh(m_\pi (T-x_0)) \,. \label{e:prop}
\end{equation}
This is confirmed by the measured data shown in Fig.~\ref{f:prop}, where
a fit of Eq.~\ref{e:prop} to the data with sufficient distance from the
boundaries is shown.

As can be seen from this example, the analysis changes  due to the
new boundary conditions, but not in a dramatic way. Experience with more
complicated correlation functions will need to be gained in order to
have a more complete picture of their consequences.

\section{Conclusions}
Few of the methods discussed in this contribution are new, still
recent years have witnessed substantial progress in our ability 
to simulate QCD with light quarks on fine lattices. This is 
mainly due to an improved understanding of these methods and 
synergies gained from combining them.

The determinant splitting techniques allow for much larger 
step sizes, which can only be understood using the 
theory of symplectic integrators. The determinant splitting, along
with techniques like twisted mass reweighting, significantly
reduce fluctuations in the forces and in  
turn make higher order and force gradient integrators proposed
during the last decade profitable. And also the sophisticated
solvers for the Dirac equation have their  setup cost
easily amortized over the many solutions needed in each individual step.
Each of these components individually already brings some gain,
but it is in their combination that the full potential can be reached.
And there might be additional gains from better further
understanding possible.

In the traditional setup, lattice QCD simulations are performed
with periodic boundary conditions, where the freezing of the topological
charge causes a severe problem with ergodicity as the 
continuum limit is approached. With typical resources, simulations
with a lattice spacing below $0.05\,\text{fm}$ are not possible, 
and the problem is sufficiently generic that an algorithmic solution seems unlikely
in the near future. Changing the lattice setup by using open boundary
conditions, however, solves the issue.

Still, QCD simulations are expensive. To keep finite volume effects
under control, the size of the box needs to be scaled with the inverse
pion mass and  physical pions require a box size of about
$6\,\text{fm}$. Fine lattice spacings are currently  difficult to
reach with such a volume.

From the simulations as described in Sec.~\ref{s:sim} but at the physical
pion mass, we can now estimate
the order of magnitude of the computer resources needed to repeat the 
simulation on a sufficiently large volume with spatial extent
$L=6\,\text{fm}$ and a finer lattice. At
$a=0.09\,\text{fm}$, we estimate $\tau_\text{int}(\overline E)\sim
20$ and we assume that this is a one of the slowest observables. If we 
want a run length of at least $100\times\tau_\text{int}(\overline E)$,
     we arrive at a cost estimate of such a simulation
     \begin{equation}
     C =
     3\,\text{Tflops}\times\text{years}\times\left(\frac{a}{0.09\,\text{fm}}\right)^{-7}
     \,.
     \end{equation}
     This means that a $a=0.045\,\text{fm}$ lattice still requires
     $400\,\text{Tflops}\times\text{years}$. Hopefully, further
     progress will reduce this number.

     \acknowledgments
It is a pleasure to thank G.~Engel, M.~L\"uscher, M.~Marinkovic,
   R.~Sommer and F.~Virotta for collaboration and many discussions
   on issues presented here. Also correspondence with B.~Leder and
   U.~Wenger is thankfully acknowledged.

\providecommand{\href}[2]{#2}\begingroup\raggedright\endgroup

\end{document}